\documentstyle[12pt,aps]{revtex}

\begin{document}

\hsize = 6.5in
\widetext
\draft
\tighten
\topmargin-48pt

\preprint{EFUAZ FT-98-55-REV}

\title{The Bargmann-Wigner Formalism for Spin 2\thanks{Submitted to
``Journal of Physics A".}}

\author{{\bf Valeri V. Dvoeglazov}}

\address{
Escuela de F\'{\i}sica, Universidad Aut\'onoma de Zacatecas \\
Apartado Postal C-580, Zacatecas 98068, ZAC., M\'exico\\
Internet address:  valeri@cantera.reduaz.mx\\
URL: http://cantera.reduaz.mx/\~\ valeri/valeri.htm
}

\date{January, 1998; revised: April 1998}

\maketitle

\begin{abstract}
We proceed to derive equations for the symmetric tensor
of the second rank on the basis of the Bargmann-Wigner formalism
in a straightforward way. The symmetric multispinor of the fourth
rank is used. It is constructed out of the Dirac 4-spinors.
Due to serious problems with the interpretation of
the results obtained on using the standard procedure we generalize
it and obtain the spin-2 relativistic equations, which are consistent
with the previous one. The importance of the 4-vector field (and its gauge
part) is pointed out.
\end{abstract}

\pacs{PACS number: 03.65.Pm , 04.50.+h , 11.30.Cp}

The general scheme for derivation of higher-spin equations
was given in~\cite{BW}. A field of rest mass $m$ and spin $j \geq {1\over
2}$ is represented by a completely symmetric multispinor of rank $2j$.
The particular cases $j=1$ and $j={3\over 2}$ were given in the
textbooks, e.~g., ref.~\cite{Lurie}. The spin-2 case can also be of some
interest because it is generally believed that the essential features of
the gravitational field are  obtained from transverse components of the
$(2,0)\oplus (0,2)$  representation of the Lorentz group. Nevertheless,
questions of the redandant components of the higher-spin relativistic
equations are not yet understood in detail~\cite{Kirch}.

In the first section we use the commonly-accepted procedure
for the derivation  of higher-spin equations.

\section{Standard Formalism}

We begin with the equations for the 4-rank symmetric spinor:
\begin{mathletters}
\begin{eqnarray}
\left [ i\gamma^\mu \partial_\mu - m \right ]_{\alpha\alpha^\prime}
\Psi_{\alpha^\prime \beta\gamma\delta} &=& 0\, ,\\
\left [ i\gamma^\mu \partial_\mu - m \right ]_{\beta\beta^\prime}
\Psi_{\alpha\beta^\prime \gamma\delta} &=& 0\, ,\\
\left [ i\gamma^\mu \partial_\mu - m \right ]_{\gamma\gamma^\prime}
\Psi_{\alpha\beta\gamma^\prime \delta} &=& 0\, ,\\
\left [ i\gamma^\mu \partial_\mu - m \right ]_{\delta\delta^\prime}
\Psi_{\alpha\beta\gamma\delta^\prime} &=& 0\, .
\end{eqnarray} \end{mathletters}
The massless limit (if one needs) should be taken in the end of all
calculations.

We proceed expanding the field function in the set of symmetric matrices
(as in the spin-1 case, cf.~ref.~[4a]). In the beginning let us use the
first two indices:\footnote{The matrix $R$ can be related to the
$CP$ operation in the $(1/2,0)\oplus (0,1/2)$ representation.}
\begin{equation} \Psi_{\{\alpha\beta\}\gamma\delta} =
(\gamma_\mu R)_{\alpha\beta} \Psi^\mu_{\gamma\delta}
+(\sigma_{\mu\nu} R)_{\alpha\beta} \Psi^{\mu\nu}_{\gamma\delta}\, .
\end{equation}
We would like to write
the corresponding equations for functions $\Psi^\mu_{\gamma\delta}$
and $\Psi^{\mu\nu}_{\gamma\delta}$ in the form:
\begin{mathletters}
\begin{eqnarray}
&&{2\over m} \partial_\mu \Psi^{\mu\nu}_{\gamma\delta} = -
\Psi^\nu_{\gamma\delta}\, , \label{p1}\\
&&\Psi^{\mu\nu}_{\gamma\delta} = {1\over 2m}
\left [ \partial^\mu \Psi^\nu_{\gamma\delta} - \partial^\nu
\Psi^\mu_{\gamma\delta} \right ]\, \label{p2}.
\end{eqnarray} \end{mathletters}
Constraints $(1/m) \partial_\mu \Psi^\mu_{\gamma\delta} =0$
and $(1/m) \epsilon^{\mu\nu}_{\quad\alpha\beta}\, \partial_\mu
\Psi^{\alpha\beta}_{\gamma\delta} = 0$ can be regarded as a consequence of
Eqs.  (\ref{p1},\ref{p2}).

Next, we present the vector-spinor and tensor-spinor functions as
\begin{mathletters}
\begin{eqnarray}
&&\Psi^\mu_{\{\gamma\delta\}} = (\gamma^\kappa R)_{\gamma\delta}
G_{\kappa}^{\quad \mu} +(\sigma^{\kappa\tau} R )_{\gamma\delta}
F_{\kappa\tau}^{\quad \mu} \, ,\\
&&\Psi^{\mu\nu}_{\{\gamma\delta\}} = (\gamma^\kappa R)_{\gamma\delta}
T_{\kappa}^{\quad \mu\nu} +(\sigma^{\kappa\tau} R )_{\gamma\delta}
R_{\kappa\tau}^{\quad \mu\nu} \, ,
\end{eqnarray}
\end{mathletters}
i.~e.,  using the symmetric matrix coefficients in indices $\gamma$ and
$\delta$. Hence, the total function is
\begin{eqnarray}
\lefteqn{\Psi_{\{\alpha\beta\}\{\gamma\delta\}}
= (\gamma_\mu R)_{\alpha\beta} (\gamma^\kappa R)_{\gamma\delta}
G_\kappa^{\quad \mu} + (\gamma_\mu R)_{\alpha\beta} (\sigma^{\kappa\tau}
R)_{\gamma\delta} F_{\kappa\tau}^{\quad \mu} + } \nonumber\\
&+& (\sigma_{\mu\nu} R)_{\alpha\beta} (\gamma^\kappa R)_{\gamma\delta}
T_\kappa^{\quad \mu\nu} + (\sigma_{\mu\nu} R)_{\alpha\beta}
(\sigma^{\kappa\tau} R)_{\gamma\delta} R_{\kappa\tau}^{\quad\mu\nu} \, ;
\end{eqnarray}
and the resulting tensor equations are:
\begin{mathletters}
\begin{eqnarray}
&&{2\over m} \partial_\mu T_\kappa^{\quad \mu\nu} =
-G_{\kappa}^{\quad\nu}\, ,\\
&&{2\over m} \partial_\mu R_{\kappa\tau}^{\quad \mu\nu} =
-F_{\kappa\tau}^{\quad\nu}\, ,\\
&& T_{\kappa}^{\quad \mu\nu} = {1\over 2m} \left [
\partial^\mu G_{\kappa}^{\quad\nu}
- \partial^\nu G_{\kappa}^{\quad \mu} \right ] \, ,\\
&& R_{\kappa\tau}^{\quad \mu\nu} = {1\over 2m} \left [
\partial^\mu F_{\kappa\tau}^{\quad\nu}
- \partial^\nu F_{\kappa\tau}^{\quad \mu} \right ] \, .
\end{eqnarray} \end{mathletters}
The constraints are re-written to
\begin{mathletters}
\begin{eqnarray}
&&{1\over m} \partial_\mu G_\kappa^{\quad\mu} = 0\, ,\quad
{1\over m} \partial_\mu F_{\kappa\tau}^{\quad\mu} =0\, ,\\
&& {1\over m} \epsilon_{\alpha\beta\nu\mu} \partial^\alpha
T_\kappa^{\quad\beta\nu} = 0\, ,\quad
{1\over m} \epsilon_{\alpha\beta\nu\mu} \partial^\alpha
R_{\kappa\tau}^{\quad\beta\nu} = 0\, .
\end{eqnarray}
\end{mathletters}
However, we need to make symmetrization over these two sets
of indices $\{ \alpha\beta \}$ and $\{\gamma\delta \}$. The total
symmetry can be ensured if one contracts the function $\Psi_{\{\alpha\beta
\} \{\gamma \delta \}}$ with {\it antisymmetric} matrices
$R^{-1}_{\beta\gamma}$, $(R^{-1} \gamma^5 )_{\beta\gamma}$ and
$(R^{-1} \gamma^5 \gamma^\lambda )_{\beta\gamma}$ and equate
all these contractions to zero (similar to the $j=3/2$ case
considered in ref.~\cite[p. 44]{Lurie}. We obtain
additional constraints on the tensor field functions:
\begin{mathletters}
\begin{eqnarray}
&& G_\mu^{\quad\mu}=0\, , \quad G_{[\kappa \, \mu ]}  = 0\, , \quad
G^{\kappa\mu} = {1\over 2} g^{\kappa\mu} G_\nu^{\quad\nu}\, ,
\label{b1}\\
&&F_{\kappa\mu}^{\quad\mu} = F_{\mu\kappa}^{\quad\mu} = 0\, , \quad
\epsilon^{\kappa\tau\mu\nu} F_{\kappa\tau,\mu} = 0\, ,\\
&& T^{\mu}_{\quad\mu\kappa} =
T^{\mu}_{\quad\kappa\mu} = 0\, ,\quad
\epsilon^{\kappa\tau\mu\nu} T_{\kappa,\tau\mu} = 0\, ,\\
&& F^{\kappa\tau,\mu} = T^{\mu,\kappa\tau}\, ,\quad
\epsilon^{\kappa\tau\mu\lambda} (F_{\kappa\tau,\mu} +
T_{\kappa,\tau\mu})=0\, ,\\
&& R_{\kappa\nu}^{\quad \mu\nu}
= R_{\nu\kappa}^{\quad  \mu\nu} = R_{\kappa\nu}^{\quad\nu\mu}
= R_{\nu\kappa}^{\quad\nu\mu}
= R_{\mu\nu}^{\quad  \mu\nu} = 0\, , \\
&& \epsilon^{\mu\nu\alpha\beta} (g_{\beta\kappa} R_{\mu\tau,
\nu\alpha} - g_{\beta\tau} R_{\nu\alpha,\mu\kappa} ) = 0\, \quad
\epsilon^{\kappa\tau\mu\nu} R_{\kappa\tau,\mu\nu} = 0\, .\label{f1}
\end{eqnarray} \end{mathletters}
Thus, we  encountered with
the known difficulty of the theory for spin-2 particles in
the Minkowski space.
We explicitly showed that all field functions become to be equal to zero.
Such a situation cannot be considered as a satisfactory one (because it
does not give us any physical information) and can be corrected in several
ways.\footnote{The reader can compare our results of this Section with
those of G. Marques and D. Spehler, Mod. Phys. Lett. A13 (1998) 553-569.
I became aware about their consideration from
Dr. D. V. Ahluwalia (personal communications, May 5, 1998) after
completing the first version of this paper. I consider their discussion
of the standard formalism in the Sections I and II, as insufficient.}

\section{Generalized Formalism}

We shall modify the formalism in the spirit of  ref.~[4b].
The field function (2) is now presented as
\begin{equation}
\Psi_{\{\alpha\beta\}\gamma\delta} =
\alpha_1 (\gamma_\mu R)_{\alpha\beta} \Psi^\mu_{\gamma\delta} +
\alpha_2 (\sigma_{\mu\nu} R)_{\alpha\beta} \Psi^{\mu\nu}_{\gamma\delta}
+\alpha_3 (\gamma^5 \sigma_{\mu\nu} R)_{\alpha\beta}
\widetilde \Psi^{\mu\nu}_{\gamma\delta}\, ,
\end{equation}
with
\begin{mathletters}
\begin{eqnarray}
&&\Psi^\mu_{\{\gamma\delta\}} = \beta_1 (\gamma^\kappa R)_{\gamma\delta}
G_\kappa^{\quad\mu} + \beta_2 (\sigma^{\kappa\tau} R)_{\gamma\delta}
F_{\kappa\tau}^{\quad\mu} +\beta_3 (\gamma^5 \sigma^{\kappa\tau}
R)_{\gamma\delta} \widetilde F_{\kappa\tau}^{\quad\mu} \, ,\\
&&\Psi^{\mu\nu}_{\{\gamma\delta\}} =\beta_4 (\gamma^\kappa
R)_{\gamma\delta} T_\kappa^{\quad\mu\nu} + \beta_5 (\sigma^{\kappa\tau}
R)_{\gamma\delta} R_{\kappa\tau}^{\quad\mu\nu} +\beta_6 (\gamma^5
\sigma^{\kappa\tau} R)_{\gamma\delta}
\widetilde R_{\kappa\tau}^{\quad\mu\nu} \, ,\\
&&\widetilde \Psi^{\mu\nu}_{\{\gamma\delta\}} =\beta_7 (\gamma^\kappa
R)_{\gamma\delta} \widetilde T_\kappa^{\quad\mu\nu} + \beta_8
(\sigma^{\kappa\tau} R)_{\gamma\delta}
\widetilde D_{\kappa\tau}^{\quad\mu\nu}
+\beta_9 (\gamma^5 \sigma^{\kappa\tau} R)_{\gamma\delta}
D_{\kappa\tau}^{\quad\mu\nu} \, .
\end{eqnarray}
\end{mathletters}
Hence, the function $\Psi_{\{\alpha\beta\}\{\gamma\delta\}}$
can be expressed as a sum of nine terms:
\begin{eqnarray}
&&\Psi_{\{\alpha\beta\}\{\gamma\delta\}} =
\alpha_1 \beta_1 (\gamma_\mu R)_{\alpha\beta} (\gamma^\kappa
R)_{\gamma\delta} G_\kappa^{\quad\mu} +\alpha_1 \beta_2
(\gamma_\mu R)_{\alpha\beta} (\sigma^{\kappa\tau} R)_{\gamma\delta}
F_{\kappa\tau}^{\quad\mu} + \nonumber\\
&+&\alpha_1 \beta_3 (\gamma_\mu R)_{\alpha\beta}
(\gamma^5 \sigma^{\kappa\tau} R)_{\gamma\delta} \widetilde
F_{\kappa\tau}^{\quad\mu} +
+ \alpha_2 \beta_4 (\sigma_{\mu\nu}
R)_{\alpha\beta} (\gamma^\kappa R)_{\gamma\delta} T_\kappa^{\quad\mu\nu}
+\nonumber\\
&+&\alpha_2 \beta_5 (\sigma_{\mu\nu} R)_{\alpha\beta} (\sigma^{\kappa\tau}
R)_{\gamma\delta} R_{\kappa\tau}^{\quad \mu\nu}
+ \alpha_2
\beta_6 (\sigma_{\mu\nu} R)_{\alpha\beta} (\gamma^5 \sigma^{\kappa\tau}
R)_{\gamma\delta} \widetilde R_{\kappa\tau}^{\quad\mu\nu} +\nonumber\\
&+&\alpha_3 \beta_7 (\gamma^5 \sigma_{\mu\nu} R)_{\alpha\beta}
(\gamma^\kappa R)_{\gamma\delta} \widetilde
T_\kappa^{\quad\mu\nu}+
\alpha_3 \beta_8 (\gamma^5
\sigma_{\mu\nu} R)_{\alpha\beta} (\sigma^{\kappa\tau} R)_{\gamma\delta}
\widetilde D_{\kappa\tau}^{\quad\mu\nu} +\nonumber\\
&+&\alpha_3 \beta_9
(\gamma^5 \sigma_{\mu\nu} R)_{\alpha\beta} (\gamma^5 \sigma^{\kappa\tau}
R)_{\gamma\delta} D_{\kappa\tau}^{\quad \mu\nu}\, .
\label{ffn1}
\end{eqnarray}
The corresponding dynamical
equations are given by the set\footnote{All indices in this formula are
already pure vectorial and have nothing to do with
previous notation. The coefficients $\alpha_i$ and $\beta_i$
may, in general, carry some dimension.}
\begin{mathletters} \begin{eqnarray}
&& {2\alpha_2
\beta_4 \over m} \partial_\nu T_\kappa^{\quad\mu\nu} +{i\alpha_3
\beta_7 \over m} \epsilon^{\mu\nu\alpha\beta} \partial_\nu
\widetilde T_{\kappa,\alpha\beta} = \alpha_1 \beta_1
G_\kappa^{\quad\mu}\,; \label{b}\\
&&{2\alpha_2 \beta_5 \over m} \partial_\nu
R_{\kappa\tau}^{\quad\mu\nu} +{i\alpha_2 \beta_6 \over m}
\epsilon_{\alpha\beta\kappa\tau} \partial_\nu \widetilde R^{\alpha\beta,
\mu\nu} +{i\alpha_3 \beta_8 \over m}
\epsilon^{\mu\nu\alpha\beta}\partial_\nu \widetilde
D_{\kappa\tau,\alpha\beta} - \nonumber\\
&-&{\alpha_3 \beta_9 \over 2}
\epsilon^{\mu\nu\alpha\beta} \epsilon_{\lambda\delta\kappa\tau}
D^{\lambda\delta}_{\quad \alpha\beta} = \alpha_1 \beta_2
F_{\kappa\tau}^{\quad\mu} + {i\alpha_1 \beta_3 \over 2}
\epsilon_{\alpha\beta\kappa\tau} \widetilde F^{\alpha\beta,\mu}\,; \\
&& 2\alpha_2 \beta_4 T_\kappa^{\quad\mu\nu} +i\alpha_3 \beta_7
\epsilon^{\alpha\beta\mu\nu} \widetilde T_{\kappa,\alpha\beta}
=  {\alpha_1 \beta_1 \over m} (\partial^\mu G_\kappa^{\quad \nu}
- \partial^\nu G_\kappa^{\quad\mu})\,; \\
&& 2\alpha_2 \beta_5 R_{\kappa\tau}^{\quad\mu\nu} +i\alpha_3 \beta_8
\epsilon^{\alpha\beta\mu\nu} \widetilde D_{\kappa\tau,\alpha\beta}
+i\alpha_2 \beta_6 \epsilon_{\alpha\beta\kappa\tau} \widetilde
R^{\alpha\beta,\mu\nu}
- {\alpha_3 \beta_9\over 2} \epsilon^{\alpha\beta\mu\nu}
\epsilon_{\lambda\delta\kappa\tau} D^{\lambda\delta}_{\quad \alpha\beta}
= \nonumber\\
&=& {\alpha_1 \beta_2 \over m} (\partial^\mu F_{\kappa\tau}^{\quad \nu}
-\partial^\nu F_{\kappa\tau}^{\quad\mu} ) + {i\alpha_1 \beta_3 \over 2m}
\epsilon_{\alpha\beta\kappa\tau} (\partial^\mu \widetilde
F^{\alpha\beta,\nu} - \partial^\nu \widetilde F^{\alpha\beta,\mu} )\, .
\label{f}
\end{eqnarray}
\end{mathletters}
Essential constraints are:
\begin{mathletters}
\begin{eqnarray}
&&\alpha_1 \beta_1 G^\mu_{\quad\mu} = 0\, ,\quad \alpha_1
\beta_1 G_{[\kappa\mu]} = 0 \, ;  \\
&&\nonumber\\
&&2i\alpha_1 \beta_2 F_{\alpha\mu}^{\quad\mu} +
\alpha_1 \beta_3
\epsilon^{\kappa\tau\mu}_{\quad\alpha} \widetilde F_{\kappa\tau,\mu} =
0\, ;\\
&&\nonumber\\
&&2i\alpha_1 \beta_3 \widetilde F_{\alpha\mu}^{\quad\mu}
+ \alpha_1 \beta_2
\epsilon^{\kappa\tau\mu}_{\quad\alpha} F_{\kappa\tau,\mu} = 0\, ;\\
&&\nonumber\\
&& 2i\alpha_2 \beta_4 T^{\mu}_{\quad\mu\alpha} -
 \alpha_3 \beta_{7}
\epsilon^{\kappa\tau\mu}_{\quad\alpha} \widetilde T_{\kappa,\tau\mu}
= 0\, ;\\
&&\nonumber\\
&& 2i\alpha_3 \beta_{7} \widetilde
T^{\mu}_{\quad\mu\alpha} -
\alpha_2 \beta_4 \epsilon^{\kappa\tau\mu}_{\quad\alpha}
T_{\kappa,\tau\mu} = 0\, ;\\
&&\nonumber\\
&& i\epsilon^{\mu\nu\kappa\tau} \left [ \alpha_2 \beta_6 \widetilde
R_{\kappa\tau,\mu\nu} + \alpha_3 \beta_{8} \widetilde
D_{\kappa\tau,\mu\nu} \right ] + 2\alpha_2 \beta_5
R^{\mu\nu}_{\quad\mu\nu}  + 2\alpha_3
\beta_{9} D^{\mu\nu}_{\quad \mu\nu}  = 0\, ;\\
&&\nonumber\\
&& i\epsilon^{\mu\nu\kappa\tau} \left [ \alpha_2 \beta_5 R_{\kappa\tau,
\mu\nu} + \alpha_3 \beta_{9} D_{\kappa\tau, \mu\nu} \right ]
+ 2\alpha_2 \beta_6 \widetilde R^{\mu\nu}_{\quad\mu\nu}
+ 2\alpha_3 \beta_{8} \widetilde D^{\mu\nu}_{\quad\mu\nu}  =0\, ;\\
&&\nonumber\\
&& 2i \alpha_2 \beta_5 R_{\beta\mu}^{\quad\mu\alpha} + 2i\alpha_3
\beta_{9} D_{\beta\mu}^{\quad\mu\alpha} + \alpha_2 \beta_6
\epsilon^{\nu\alpha}_{\quad\lambda\beta} \widetilde
R^{\lambda\mu}_{\quad\mu\nu} +\alpha_3 \beta_{8}
\epsilon^{\nu\alpha}_{\quad\lambda\beta} \widetilde
D^{\lambda\mu}_{\quad \mu\nu} = 0\, ;\\
&&\nonumber \\
&&2i\alpha_1 \beta_2 F^{\lambda\mu}_{\quad\mu} - 2 i \alpha_2 \beta_4
T_\mu^{\quad\mu\lambda} + \alpha_1 \beta_3 \epsilon^{\kappa\tau\mu\lambda}
\widetilde F_{\kappa\tau,\mu} +\alpha_3 \beta_7
\epsilon^{\kappa\tau\mu\lambda} \widetilde T_{\kappa,\tau\mu} =0\, ;\\
&&\nonumber\\
&&2i\alpha_1 \beta_3 \widetilde F^{\lambda\mu}_{\quad\mu} - 2 i \alpha_3
\beta_7 \widetilde T_\mu^{\quad\mu\lambda} + \alpha_1 \beta_2
\epsilon^{\kappa\tau\mu\lambda} F_{\kappa\tau,\mu} +\alpha_2
\beta_4 \epsilon^{\kappa\tau\mu\lambda}  T_{\kappa,\tau\mu} =0\, ;\\
&&\nonumber\\
&&\alpha_1 \beta_1 (2G^\lambda_{\quad\alpha} - g^\lambda_{\quad\alpha}
G^\mu_{\quad\mu} ) - 2\alpha_2 \beta_5 (2R^{\lambda\mu}_{\quad\mu\alpha}
+2R_{\alpha\mu}^{\quad\mu\lambda} + g^\lambda_{\quad\alpha}
R^{\mu\nu}_{\quad\mu\nu}) +\nonumber\\
&+& 2\alpha_3 \beta_9
(2D^{\lambda\mu}_{\quad\mu\alpha} + 2D_{\alpha\mu}^{\quad\mu\lambda}
+g^\lambda_{\quad\alpha} D^{\mu\nu}_{\quad\mu\nu})+
2i\alpha_3 \beta_8 (\epsilon_{\kappa\alpha}^{\quad\mu\nu}
\widetilde D^{\kappa\lambda}_{\quad\mu\nu} -
\epsilon^{\kappa\tau\mu\lambda} \widetilde D_{\kappa\tau,\mu\alpha}) -
\nonumber\\
&-& 2i\alpha_2 \beta_6 (\epsilon_{\kappa\alpha}^{\quad \mu\nu}
\widetilde R^{\kappa\lambda}_{\quad\mu\nu} -
\epsilon^{\kappa\tau\mu\lambda} \widetilde R_{\kappa\tau,\mu\alpha})
= 0\, ; \\
&&\nonumber\\
&& 2\alpha_3 \beta_8 (2\widetilde D^{\lambda\mu}_{\quad\mu\alpha} + 2
\widetilde D_{\alpha\mu}^{\quad\mu\lambda} +g^\lambda_{\quad\alpha}
\widetilde D^{\mu\nu}_{\quad\mu\nu}) - 2\alpha_2 \beta_6 (2\widetilde
R^{\lambda\mu}_{\quad\mu\alpha} +2 \widetilde
R_{\alpha\mu}^{\quad\mu\lambda} + \nonumber\\
&+& g^\lambda_{\quad\alpha} \widetilde
R^{\mu\nu}_{\quad\mu\nu}) +
+ 2i\alpha_3 \beta_9 (\epsilon_{\kappa\alpha}^{\quad\mu\nu}
D^{\kappa\lambda}_{\quad\mu\nu}  - \epsilon^{\kappa\tau\mu\lambda}
D_{\kappa\tau,\mu\alpha} ) -\nonumber\\
&-& 2i\alpha_2 \beta_5
(\epsilon_{\kappa\alpha}^{\quad\mu\nu} R^{\kappa\lambda}_{\quad\mu\nu}
- \epsilon^{\kappa\tau\mu\lambda} R_{\kappa\tau,\mu\alpha} ) =0\, ;\\
&&\nonumber\\
&&\alpha_1 \beta_2 (F^{\alpha\beta,\lambda} - 2F^{\beta\lambda,\alpha}
+ F^{\beta\mu}_{\quad\mu}\, g^{\lambda\alpha} - F^{\alpha\mu}_{\quad\mu}
\, g^{\lambda\beta} ) - \nonumber\\
&-&\alpha_2 \beta_4 (T^{\lambda,\alpha\beta}
-2T^{\beta,\lambda\alpha} + T_\mu^{\quad\mu\alpha} g^{\lambda\beta} -
T_\mu^{\quad\mu\beta} g^{\lambda\alpha} ) +\nonumber\\
&+&{i\over 2} \alpha_1 \beta_3 (\epsilon^{\kappa\tau\alpha\beta}
\widetilde F_{\kappa\tau}^{\quad\lambda} +
2\epsilon^{\lambda\kappa\alpha\beta} \widetilde F_{\kappa\mu}^{\quad\mu} +
2 \epsilon^{\mu\kappa\alpha\beta} \widetilde F^\lambda_{\quad\kappa,\mu})
-\nonumber\\
&-& {i\over 2} \alpha_3 \beta_7 ( \epsilon^{\mu\nu\alpha\beta} \widetilde
T^{\lambda}_{\quad\mu\nu} +2 \epsilon^{\nu\lambda\alpha\beta} \widetilde
T^\mu_{\quad\mu\nu} +2 \epsilon^{\mu\kappa\alpha\beta} \widetilde
T_{\kappa,\mu}^{\quad\lambda} ) =0\, .
\end{eqnarray}
\end{mathletters}
They are  the results of contractions of the field function (\ref{ffn1})
with three antisymmetric matrices, as above. Furthermore,
one should recover the relations (\ref{b1}-\ref{f1}) in the particular
case when $\alpha_3 = \beta_3 =\beta_6 = \beta_9 = 0$ and
$\alpha_1 = \alpha_2 = \beta_1 =\beta_2 =\beta_4
=\beta_5 = \beta_7 =\beta_8 =1$.

As a discussion we note that in such a framework we already have physical
content because only certain combinations of field functions
would be equal to zero. In general, the fields
$F_{\kappa\tau}^{\quad\mu}$, $\widetilde F_{\kappa\tau}^{\quad\mu}$,
$T_{\kappa}^{\quad\mu\nu}$, $\widetilde T_{\kappa}^{\quad\mu\nu}$, and
$R_{\kappa\tau}^{\quad\mu\nu}$,  $\widetilde
R_{\kappa\tau}^{\quad\mu\nu}$, $D_{\kappa\tau}^{\quad\mu\nu}$, $\widetilde
D_{\kappa\tau}^{\quad\mu\nu}$ can  correspond to different physical states
and the equations above describe oscillations one state to another.

Furthermore, from the set of equations (\ref{b}-\ref{f}) one
obtains the {\it second}-order equation for symmetric traceless tensor of
the second rank ($\alpha_1 \neq 0$, $\beta_1 \neq 0$):
\begin{equation} {1\over m^2} \left [\partial_\nu
\partial^\mu G_\kappa^{\quad \nu} - \partial_\nu \partial^\nu
G_\kappa^{\quad\mu} \right ] =  G_\kappa^{\quad \mu}\, .
\end{equation}
After the contraction in indices $\kappa$ and $\mu$ this equation is
reduced to the set
\begin{mathletters}
\begin{eqnarray}
&&\partial_\mu G_{\quad\kappa}^{\mu} = F_\kappa\,  \\
&&{1\over m^2} \partial_\kappa F^\kappa = 0\, ,
\end{eqnarray}
\end{mathletters}
i.~e.,  to the equations connecting the analogue of the energy-momentum
tensor and the analogue of the 4-vector potential. As we showed in our
recent work~\cite{Dvo97} the longitudinal potential is perfectly suitable
for construction of electromagnetism (see also recent works on the notoph
and notivarg concept~\cite{Tybor}).  Moreover, according to the Weinberg
theorem~\cite{Weinberg} for massless particles it is the gauge part of the
4-vector potential which is the physical field. The case, when the
longitudinal potential is equated to zero, can be considered as a
particular case only.  This case may be relevant to some physical
situation but hardly to be considered as a fundamental one.

Further investigations may provide additional foundations to
``surprising" similarities of gravitational and electromagnetic
equations in the low-velocity limit, refs.~\cite{Wein2,Jef}.

\acknowledgements
I acknowledge discussions with Profs. D. V. Ahluwalia, A. E. Chubykalo,
S. Esposito, Y.~S. Kim, A.~F.  Pashkov, G. Quznetzov and R. Santilli.

I am grateful to Zacatecas University, M\'exico, for the
professorship.  This work has been partly supported by the Mexican
Sistema Nacional de Investigadores, by the CONACyT, M\'exico under the
research project 0270P-E.

\end{document}